\documentclass{PoS}
\def\sigv{\langle\sigma v\rangle}

\title{WIMP Dark Matter and the First Stars: \\ a critical overview}

\ShortTitle{PopIII \& Dark Matter}

\author{\speaker{Fabio Iocco}\\
        {\it Marie-Curie} fellow\footnote{Supported
by European Community research program FP7/2007/2013 
within the framework of convention \#235878. } at\\
        Institut d'Astrophysique de Paris, 
        UMR 7095-CNRS, Universit\'e Pierre et Marie Curie, \\
        98 bis Boulevard Arago 75014, Paris, France\\
        E-mail: \email{iocco@iap.fr}}

\abstract{
If Dark Matter (DM) is composed by 
Weakly Interacting Massive Particles,
its annihilation in the halos harboring the
earliest star formation episode may strongly
influence the first generation of stars (Population III).
Whereas DM annihilation at early stages of
gas  collapse does not dramatically affect the
properties of the cloud, the formation
of a hydrostatic object (protostar) and its evolution
toward the main sequence may be delayed.
This process involves DM concentrated in the
center of the halo by gravitational drag, and
no consensus is yet reached over whether this can 
push the initial mass of Population III to higher masses.
DM can also be captured through scattering over the baryons
in a dense object, onto or very close to the Main Sequence.
This mechanism can affect formed stars and in
principle prolonge their lifetimes.
The strength of both mechanisms depends upon 
several environmental conditions and on DM parameters;
such spread in the parameter space leads to
very different scenarios for the observables in the
Population.
Here I summarize the state of the art in modelling and
observational expectations, eventually highlighting
the most critical assumptions and sources of uncertainty.}

\FullConference{Cosmic Radiation Fields: Sources in the early Universe - CRF2010,\\
		November 9-12, 2010\\
		Desy Germany}

\begin{document}

In Lambda cold dark matter ($\Lambda$CDM) cosmology, most of the matter in 
the Universe is in the form of one or more electromagnetically undetected 
species of particles whose only known influence is gravitational.  If dark 
matter is composed of WIMPs that self-annihilate, such annihilations 
occur in the particularly high central densities of collapsing primordial 
gas clouds, and the energy released may affect the first 
stars (Population III) by two different mechanisms.
\section{Gravitational accretion of DM}
DM self-annihilates and injects energy everywhere inside a primordial 
star-forming halo at a rate that depends on the local DM density and 
the opacity of the gas to the high-energy primaries produced in the 
annihilation \cite{Bertone:2004pz,Ascasibar07}.
The deposited energy per unit volume, per unit time
can be written as:
\begin{equation}
\frac{dL}{dV}(r)=n^2_{DM}(r)\sigv m_{DM}\times\kappa (r)
\label{Lumin}
\end{equation}
with $n_{DM}(r)$ being the DM number density at a given radius,
 $\kappa (r)$ the fraction of  energy deposited in the gas from the primary
shower induced by DM annihilation, $\sigv$ is the velocity
averaged self-annihilation rate of DM; the formula
clearly showing that the effects are going to be strongest in the
central regions of the halo, where gas (and DM) densities are typically highest.

The effects of 
this process are more dramatic in high-z minihalos than in the local Universe;
this is due to the peculiarities of PopIII star formation,
which is believed to take place in the very center of the DM halo,
following a very smooth collapse of the gas, consequence of
the absence of strong coolants, see e.g.  \cite{Bromm:2009uk}.
This favors the build-up of a central, massive gaseous object,
which can gravitationally drag the collisionless matter, 
permitting the formation of very high densities of DM, \cite{Spolyar:2007qv}.  
In the latter, using a 
semianalytical model without feedback on primordial gas chemistry,  
the authors found that for a wide range of DM parameters the gas cloud always 
enters a stage in which energy injection due to DM annihilation equals the 
feeble H$_2$ cooling in the innermost region of the cloud.  
This occurs 
before a hydrostatic object forms there, when central 
densities reach $n_c\sim$ 10$^{12}$ cm$^3$ (the actual value 
depending on DM mass, $\sigv$, and primary annihilation channels). 
They speculated that the gas cloud could actually halt its collapse and form an object 
powered by DM annihilation, which they refer to as a ``{\it dark star}''.

Most recently, full 1-D hydrodynamical simulations have been performed
which properly model gas chemistry, the coupling of DM annihilations to 
chemistry, and the response of DM to the variation of gravitational 
potential induced by baryonic collapse (which is modeled through the
so-called  ``adiabatic''  contraction approximation). 

By taking into account the feedback of DM annihilations on 
gas chemistry, we have shown in \cite{Ripa1} that although they {\it 
do} modify the temperature and ionization state of the cloud, these 
effects are however vastly mitigated by their feedback. The high energy shower 
is injected at the center of the cloud by annihilating DM (where annihilation 
rates are greatest and couple most efficiently to the gas) and it {\it does} heat 
the halo core and its surrounding regions.  
However, we find that the large ionized
fractions induced by the annihilations impact the chemistry, which in turn 
regulates the temperature, eventually without dramatic effects on the stability of the 
cloud or its Jeans mass. 

Following the collapse of the cloud beyond the time when DM annihilation 
heating equals chemical cooling (for a variety of astrophysical 
and DM parameters) we find that the collapse does {\bf not} halt, and a ``dark
star'' does {\bf not} form, at least at this stage.  Simulations capable of following
the formation of a hydrostatic core and DM annihilations therein (even better,
doing so in 3D) are needed to determine the nature of the resulting proto-star. 
We have anyways assumed that eventually a hydrostatic object
powered by annihilation of the local (object-embedded) DM environment
can form, and we have wondered which may be its properties.
We have modeled a hydrostatic object, powered by energy from 
the annihilations of DM embedded in the cloud, the ``local bath''.
The latter is assumed to be following the gravitational contraction of the gas 
using the so-called adiabatic contraction approximation \cite{Iocco:2008rb}.
We find that the equilibrium of such objects is unstable, and that DM 
annihilation can only delay collapse for times that are short compared to 
contraction timescales -$\mathcal{O}$(10$^4$-10$^5$yr)-,
which again suggests that the annihilation of 
gravitationally contracting DM cannot give rise to any stable or long-lived 
phase.  This is in contrast with results achieved by \cite{Freese:2008wh}, 
that observe equilibrium between DM gravitational accretion and annihilation,
and gravitational collapse of gas, over times of $\mathcal{O}$(10$^6$yr); by including
gas accretion on the hydrostatic, stable object, they observe the formation
of extremely massive objects, up to 10$^5$M$_\odot$, before they reach the 
nuclear burning and collapse to Black Holes.

It is worth stressing that the process described so far can take place
only once during the life of a celestial object: it is intrinsically
related to the contraction phase of a {\it pre} and then {\it proto}
stellar object, and its contraction toward the hydrostatic equilibrium
and then the Main Sequence. Such process
is characteristic of a metal-free, smooth collapse without fragmentation during early stages.
This process is usually (and almost univocally) associated with Population III stars,
and alien to galactic star formation.

\section{Scattering accretion of DM}
While collapsing, a ``DM-affected proto-PopIII'' moves out of the Hayashi track.
At this point the object -either with a humongous mass built-up
under the effect of DM sustaining or with a ``traditional'' PopIII mass- ignites
nuclear reactions as a consequence of structure contraction and heating;
however, the star is still embedded in a very high DM bath,
 {\it external} to the star itself. 
No earlier than this point in proto-stellar evolution, and
if the cross-section between baryons and WIMPs is high enough,  
capture of DM in the star via a {\it scattering} process becomes efficient. 
That is, DM captured onto the star by scattering
will thermalize and ``sink'' in an equilibrium configuration
in a small region of the stellar core itself  within short timescales. 
The annihilation of such concentrated
distribution can indeed power the star entirely (depending on parameters), 
as found in \cite{Iocco:2008xb}.  
The most dramatic effect of such ``DM burning'' is that 
the star either halts its contraction before hydrogen burning or later causes
{\it H} to burn at a reduced rate because DM energy release supports the star against 
further collapse \cite{Iocco:2008rb, CaptEvol}. 
This implies that the duration 
of the main sequence is prolonged until most of the hydrogen in the core is 
converted into helium at such slow rate,
 and that the subsequent chemical evolution (and thus
aging) of the star are delayed. A DM-burning star's evolution is frozen as long
as DM capture can proceed at the rate necessary to power entirely the stellar 
luminosity.  The formal details of scattering and gravitational accretion 
are described in the original literature \cite{Iocco:2008rb, CaptEvol}, 
and summarized in previous proceedings \cite{DSprocs} and in \cite{Zackrisson:2010jd}.  
It is worth stressing that while gravitational accretion depends only on the 
self-annihilation of the DM density field in the core of the cloud, scattering 
accretion involves DM particles originally outside of it, and relies on the 
existence of a sizeable scattering cross section between baryons and WIMPs 
to enable DM capture within the star.  On the other hand, the 
scattering-driven annihilation process is not as intrinsically unstable as the 
gravitational-accretion process, and can apply to galactic stars, provided that
the environment DM density is high enough. This is possibly the case of
regions around the central Black Hole, see \cite{Scott:2008ns}.

\section{Detection prospects}
\begin{center}
\begin{figure}
  \includegraphics[height=.2\textheight]{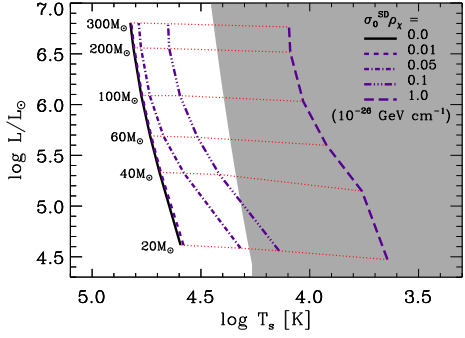}
  \caption{The HR diagram of massive, metal free stars of several masses that
  are influenced by scatter-captured DM burning, for different values of the 
  product of the spin-dependent elastic scattering cross section $\sigma_0$
  and the DM density $\rho_\chi$ in which the star is embedded.
  From Yoon, Iocco, and Akiyama `08 in \cite{CaptEvol}.}
 \label{HRplot}
\end{figure}
\end{center}

The delay of nuclear ignition in a hydrostatic protostar by the DM scattering
process can be visualized as an interrupted track toward the ZAMS in the HR 
diagram. The position of Pop III stars of different masses 
(and with different DM parameters) is shown in Fig \ref{HRplot}: 
DM-burning stars in the grey 
region are entirely supported by scattering-accreted DM annihilations as long 
as WIMPs can be replenished.  DM-burning Pop III stars (those that are {\it 
entirely} supported by energy from DM annihilation, sometimes
referred to as {\it dark stars}) are colder and larger than 
normal Pop III stars.  The nature of dark stars is critical to determine 
their observational signatures. In principle, the life-prolonging
effect spread over an entire population, coul decouple the pair-instability 
supernova (PISN) rates from star formation rate (Iocco (2009) in \cite{DSprocs}), 
especially at high redshifts. 
Other authors have studied whether 
a dark star could be directly observed by JWST; 
as a matter of fact
the possibility to detect dark stars
with current instruments (i.e. HST) has already been ruled out for reasonable models 
\cite{Zackrisson:2010jd}.  
We concluded that even JWST will be unable to 
observe directly any of these objects unless they are lensed by massive 
intervening structures along the line of sight; even then, detection would be 
difficult. 

However, the life-prolonging effect of DM burning on stars may allow several 
of these objects to be conveyed up in the first galaxies, following the
halo merger history.  This would imprint  peculiar signatures on protogalaxies, 
which would be recognizable 
in the JWST fields.  However, the number and characteristics of affected and 
detectable galaxies varies strongly with the parameters assumed for the nature
of DM particles (like the elastic scattering cross section between baryons and 
WIMPs), the size of the central cusp of the DM halo, the likelihood of star 
formation within it, and the number of Pop III stars that form in the halo. 
Such sensitivity could constitute a diagnostic tool for discriminating between 
formation scenarios; however, for now, detectability of these objects lies 
beyond our reach.

\section{Which {\it Population} effects?}
Wondering if Population III stars affected by
Dark Matter are observable, is equivalent to ask whether
these effects are really so dramatic at all.
The possibility to observe an indirect signature of ``DM burning'',
would implicitly mean that the gap between the standard and the
exotic scenario is significant, and that the whole Population III is strongly
affected by this mechanism.

On one hand our discriminant power is plagued by the fact
that we have no observational evidence for the standard Population III,
that is to say we know very little, and our expectations strongly vary with the
parameters of the "standard" PopIII model.
On the other, it seems that DM burning will little affect the final properties
of the Population, thus creating little hope for observations, at least now.

Nonetheless, it is still extremely interesting to wonder whether Dark Matter
has affected, beyond the pure gravitational or ``inert'' effects, a whole population of stars.
Whereas the motivations leading us to believe it plausible are strong (as summarized
so far) yet they rely on untested, although extremely reasonable, assumptions.
Here is a list of what skepticists should definitely consider as the most ucertain
ingredients of the models.

\subsection{Open issues}
The formation of {\it i)}: a DM dense profile (hereafter called ``cusp'', 
whereas it could also be cored, this definition referring to a density enhanced 
with respect to the original profile) is a condition for both 
the {\it gravitational} and {\it scattering} phases. 
Beside being predicted by seminanalitic models
adopting different flavors of the ``adiabatic contraction'' approximation
\cite{Spolyar:2007qv, Iocco:2008rb},
the enhancement of the DM profile due to gravitational drag of the baryons
has been observed in simulations down to the resolved scale 
($n_{gas}\lesssim$10$^{12}$\#/cm$^3$, $r\lesssim$10$^{-1}$pc),
by \cite{Natarajan:2008db}.
Whereas a still higher density is to be expected in the unresolved center,
even a plateau at the level of the one found in the innermost resolved region
would be enough for the first stages of gravitational contraction.
Which raises the issue {\it ii)}:  the alignment between object and the DM cusp.
In PopIII formation the definition of the halo ``centering'' is well posed: the gravitational 
potential being given by the collapse of only one (baryonic) object, the DM will in first
approximation overlap with the source of such radial profile. This approximation holds
until very fine tuning is needed: for the {\it gravitational} phase
the hydrostatic object of size of $\mathcal{O}$(1AU) must lie within a DM cusp with radius of
$\mathcal{O}$(10$^2$--10$^3$AU);
in the {\it scattering} phase the size of the region with the right DM density 
is of $\mathcal{O}$(10--10$^2$AU), whereas the size of the hydrostatic
object varies between one and ten solar radii $\mathcal{O}$(0.01-0.1AU).
With respect to this, it worth noticing that {\it iii)}: 3-dimensional effects 
have not yet been included when following the gravitational effects:
these might have consequences on both the aligment argument and the following issue,
that {\it iv)}: adiabatic contraction approximation is used
in order to compute the DM build-up during the {\it gravitational} phase of the
hydrostatic object. This might overestimate to some extent the DM response to baryons,
see discussion in Ripamonti et al. `10. It has been recently argued that
{\it v)}: DM in the surrouding of the hydrostatic object
can be exhausted by annihilation at such level to contribute
virtually nothing to the {\it scattering} phase, \cite{Sivertsson:2010zm}.
These results depend however quite strongly on the initial conditions of the 
DM halo and the stellar mass, and might not apply in all cases.

This list of the critical issues of the ``Dark Stars'' scenario shows that,
whereas there are very sound reasons to argue about the
possible relevance WIMP DM effects onto the Population III,
our predictive power is very low, as little do we know about these
critical issues and can only make predictions within acceptable range
of parameters. 
Second, it is reasonable to expect that the global 
effects on the whole Population
will be limited, as conditions for ``Dark Stars''
existence would be the outcome of concurrence of several
extremely favorable coincidences.
\begin{center}
{\it Note added}
\end{center}

It is to be noticed that all existing literature addressing the effects
of WIMP DM on PopulationIII stars relies on
the paradygm of single star formation. Such scenario,
widely agreed upon until very recently, has been challenged in the last
couple of years by research showing that binary or multiple stellar 
systems may  not be so unique in the early Universe 
star formation, \cite{multiplePOPIII}.
How this would affect the scenario depicted in these
proceedings is yet to be explored in details; still, 
some useful considerations may be drawn by noticing 
the following few keypoints.

The fragmentation of the halo takes place at gas densities
smaller or comparable with those of the {\it gravitational} phase
(e.g. in Turk et al. `09, $n_{gas}\sim$10$^{11}$\#/cm$^3$).
This means that {\it a)} one or more fragments
could be involved already
during the {\it gravitational} phase, and 
concerns about deviations from 
sphericity (see {\it ii)} and {\it iii)} become more
motivated, as well as those
on incorrect application of the adiabatic contraction
mechanism (see {\it iv}).
Notice however that the fragments
(proto-stars)  are concentrated in 
the DM cusp region, thus {\it b)} still 
making conditions for
the {\it scattering} phase possible.
Yet, how DM 
distribution in phase space reacts to such 
baryonic stirring, must be looked upon,
so far leaving any scenario for
DM and multiple PopIII open.

\begin{center}
{\bf Acknowlegements}
\end{center}
I would like to thank the organizers for the invitation, and my co-authors
on the subject of Dark Matter and Stars present at this meeting: 
A. Ferrara, C.E. Rydberg, E. Ripamonti,  
P. Scott, E.  Zackrisson, for fruitful discussions.

\end{document}